\documentclass[%
 reprint,
 amsmath,amssymb,
 aps,
]{revtex4-1}

\usepackage{graphicx}
\usepackage{dcolumn}
\usepackage{bm}

\begin{document}

\preprint{APS/123-QED}

\title{Direct Photon Production and Azimuthal Anisotropy at Low Transverse Momentum measured in PHENIX}


\author{Wenqing Fan for PHENIX collaboration}

\address{}

\begin{abstract}
The PHENIX experiment discovered a large excess of low-$p_{T}$ direct photons in Au+Au collisions at 200 GeV compared to reference p+p collisions, which has been attributed to thermal radiation from the medium produced in the collisions. At the same time the excess photons show a large azimuthal anisotropy, expressed as Fourier coefficients $v_{2}$ and $v_{3}$. These surprising results have not yet been fully described by theoretical models. 
We will present 
the results obtained from real photons in the electromagnetic calorimeter and photons converted on the outer shell of the Hadron Blind Detector.
PHENIX has also developed a new technique to identify conversion photons without assuming the radius where the conversion happened. This method greatly increases the available statistics and reduces systematic uncertainties.   
\end{abstract}

\maketitle


\section{Introduction}
\label{}
The main goal of colliding heavy ions is to study the the properties of Quark Gluon Plasma (QGP), and direct photons have long been considered a golden probe for that. Unlike hadronic observables, direct photons are ``color blind" probes, which means they will not interact strongly with the medium hence leave the medium unscathed. Also, direct photons are produced at almost all known or conjectured stages of the collision. Therefore, by measuring direct photons one has access to the information about the entire evolution of the colliding system.

Experimentally, one measures a time-integrated history of the emission. Once photons from hadronic decays are subtracted, the so-called direct photons remain, which can come from hard scattering, thermal radiation of the QGP, thermal radiation of the hadron gas (HG), or other sources like Bremsstrahlung. More differential measurements may be needed to help disentangle the multiple sources and separate the QGP thermal signal from the rest.

PHENIX has measured the direct photon yield \cite{adare2010enhanced} and azimuthal anisotropy from Au+Au collisions at RHIC energies. A large excess of direct photons are observed at low $p_{T}$ with strong centrality dependence of the yield. Quite surprisingly, the excess photons exhibit a large azimuthal anisotropy with respect to the event plane, quantified as the Fourier coefficient $v_{2}$ and $v_{3}$. The simultaneous observations of large yield and large azimuthal anisotropy of direct photons contradict several existing theoretical models, for instance hydrodynamic models and microscopic transport models. The source of the problem is the interplay between the build-up of the collective flow and the cooling as the system expands. The large yield indicates emissions from an early stage when the temperature is sufficiently high. Conversely, the large anisotropy implies emissions from a late stage when the collective flow is fully developed. The failure to account for the large yield and anisotropy simultaneously is called ``direct photon puzzle''.

To resolve the puzzle, a lot of new theoretical models were proposed. On the experimental side, PHENIX has made a new measurement on the azimuthal anisotropy of direct photon emissions from Au+Au collisions at 200\,GeV recorded in 2007 and 2010, as presented in~\cite{adare2016azimuthal}. We also developed a new technique to identify conversion photons without assuming the conversion radius. Applying this new technique to analyze the photon conversions at the layers of Silicon Vertex Tracker (VTX) Detector in the 2014 Au+Au dataset, we will be able to have more precise results on direct photons.

\section{Experimental Setup and Photon Measurement}
\label{}
The detectors involved in these measurements are the tracking system to measure the trajectories and momenta of charged particles, the electromagnetic calorimeter (EMCal) to determine the energy photons and electrons, and the Ring Imaging Cerenkov detector (RICH) to identify electrons (positrons). The tracking system consists of three parts: the central arm magnet, two drift chambers (DC) and the first pad chamber (PC1). DC and PC1 measure the deflection of charged particles in the axial magnetic field. The EMCal comprises of two calorimeter types: 6 sectors of lead scintillator sampling calorimeter (PbSc) and  2 sectors of lead glass Cerenkov calorimeter (PbGl). The typical energy resolution of the PbSc is $\delta E/E = 8.1\%/\sqrt{E(GeV)}\oplus2.1\%$, and that of the PbGl is $\delta E/E = 5.9\%/\sqrt{E(GeV)}\oplus0.8\%$.

\begin{figure}[htb] 
        \centering \includegraphics[width=0.8\columnwidth]{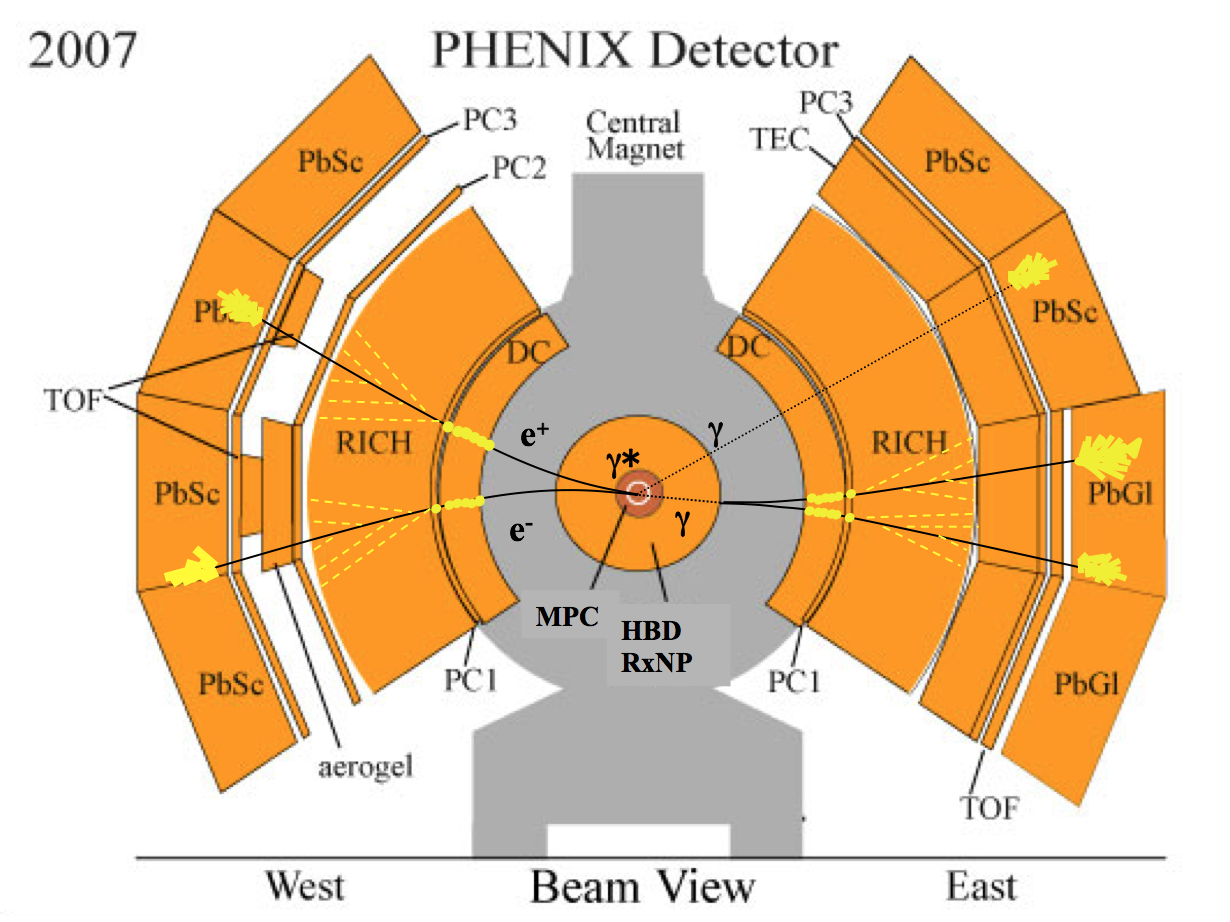}
		\caption{
        		Schematic view of the three methods for photon measurement in PHENIX. Dashed line corresponds to photon trajectory, solid curve corresponds to $e^{\pm}$ trajectory.}
		\label{fig:methods}
\end{figure}

In PHENIX, photons are measured by three different techniques, as shown in Fig.~\ref{fig:methods}. Historically, the calorimeter method was used first \cite{adare2012observation}. It measures photons via their energy deposit in the electromagnetic calorimeter (EMCal). This method has good resolution at high $p_{T}$, but suffers from the hadron contamination in the low $p_{T}$ regime. To reduce the hadron background, the internal conversion method has been developed. It measures virtual photons converting to $e^{+}$ and $e^{-}$ still in the interaction vertex; the $e^{\pm}$ pair is reconstructed by the tracking system, and identified by the RICH and the EMCal. This method provides a high purity photon sample, but it is limited in statistics due to the low internal conversion probability and reduced acceptance. The third method is the external conversion method \cite{adare2015centrality}, which measures photons that convert on some real detector material, far away from the vertex. It provides more statistics compared to internal conversion and good resolution at low $p_{T}$. The results presented here use the calorimeter method and the external conversion method.

In the external conversion method, the inclusive photon sample is provided by those $e^{\pm}$ pairs that are conversions at the readout plane of the Hadron Blind Detector (HBD), located at ~60\,cm radial distance from the collision vertex and $\sim3\%$ radiation length thick. The electron and positron from photon conversion are identified by the RICH and a matching cluster of energy E in the EMCal such that $E>0.15GeV$ and $E/p>0.5$, where $p$ is the momentum reconstructed by the central tracking system. All surviving $e^{\pm}$ tracks with $p_{T}>0.2GeV/c$ are combined to pairs, and conversion photons are identified by the invariant mass of the pairs. Due to the default assumption that all tracks originate from the collision vertex, the $e^{+}e^{-}$ from real conversions at the HBD readout plane will be mis-reconstructed with an artificial opening angle and hence a finite mass $m_{ee} \sim12MeV/c^{2}$. Conversely, if the reconstruction is performed assuming the HBD readout plane as the origin, the invariant mass is close to zero. By cutting on both masses simultaneously, a photon sample of 99$\%$ purity is obtained down to $p_{T} = 0.4\,GeV/c$. The photon identification in the calorimeter method is more straightforward. The photon candidates are EMCal clusters above a threshold energy of 0.2GeV that pass a shower shape cut as well as a charged particle veto cut by the pad chamber PC3.


\section{Results}
\label{}
In order to measure photon $v_{n}$, the event plane method is used. We measure the azimuthal distribution of photons with respect to the event plane $\Psi_n$, determined by the inner and outer Reaction Plane Detectors (RxN) \cite{stankus2011reaction}. $v_{n}$ are obtained from the coefficients of Fourier decomposition
\begin{align}
\frac{dN}{d(\phi-\Psi_{n})} \propto 1+\sum\limits_{n}2v_{n}\mbox{cos}(n(\phi-\Psi_{n}))
\end{align}

The direct photon $v_{2}^{dir}$ and $v_{3}^{dir}$ are calculated from the inclusive photon $v_{n}^{inc}$ and the (simulated) decay photon $v_{n}^{dec}$, as well the ratio $R_{\gamma}$ of the total inclusive and decay photon yields, using the formula \cite{adare2012observation}:
\begin{align}
v_{n}^{dir} = \frac{R_{\gamma}v_{n}^{inc}-v_{n}^{dec}}{R_{\gamma}-1}
\end{align}

\begin{figure}[htb] 
        \centering \includegraphics[width=1.0\columnwidth]{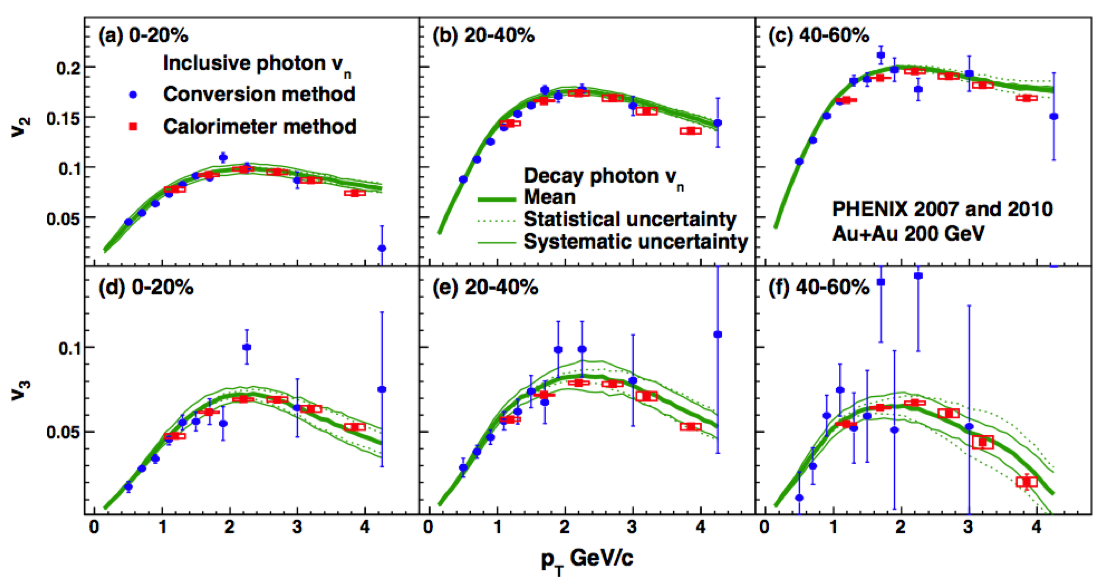}
		\caption{
        		Inclusive photon $v_{2}^{inc}$ and $v_{3}^{inc}$ at midrapidity ($|\eta|<0.35$) for Au+Au at 200GeV in different centrality bins 0-20$\%$ (a, d), 20-40$\%$ (b, e), 40-60$\%$ (c, f). The data from the external conversion method are shown as solid circles and from calorimeter method are shown as solid squares. Also shown are the calculated decay photon$v_{2}^{dec}$ and $v_{3}^{dec}$ in thick solid lines.}
		\label{fig:vn_incl}
\end{figure}

As the first step, $v_{2}^{inc}$ and $v_{3}^{inc}$ are determined for the (inclusive) conversion photon sample and for the calorimeter photon sample via event plane method (see  Fig.~\ref{fig:vn_incl}).

\begin{figure}[htb] 
        \centering \includegraphics[width=0.8\columnwidth]{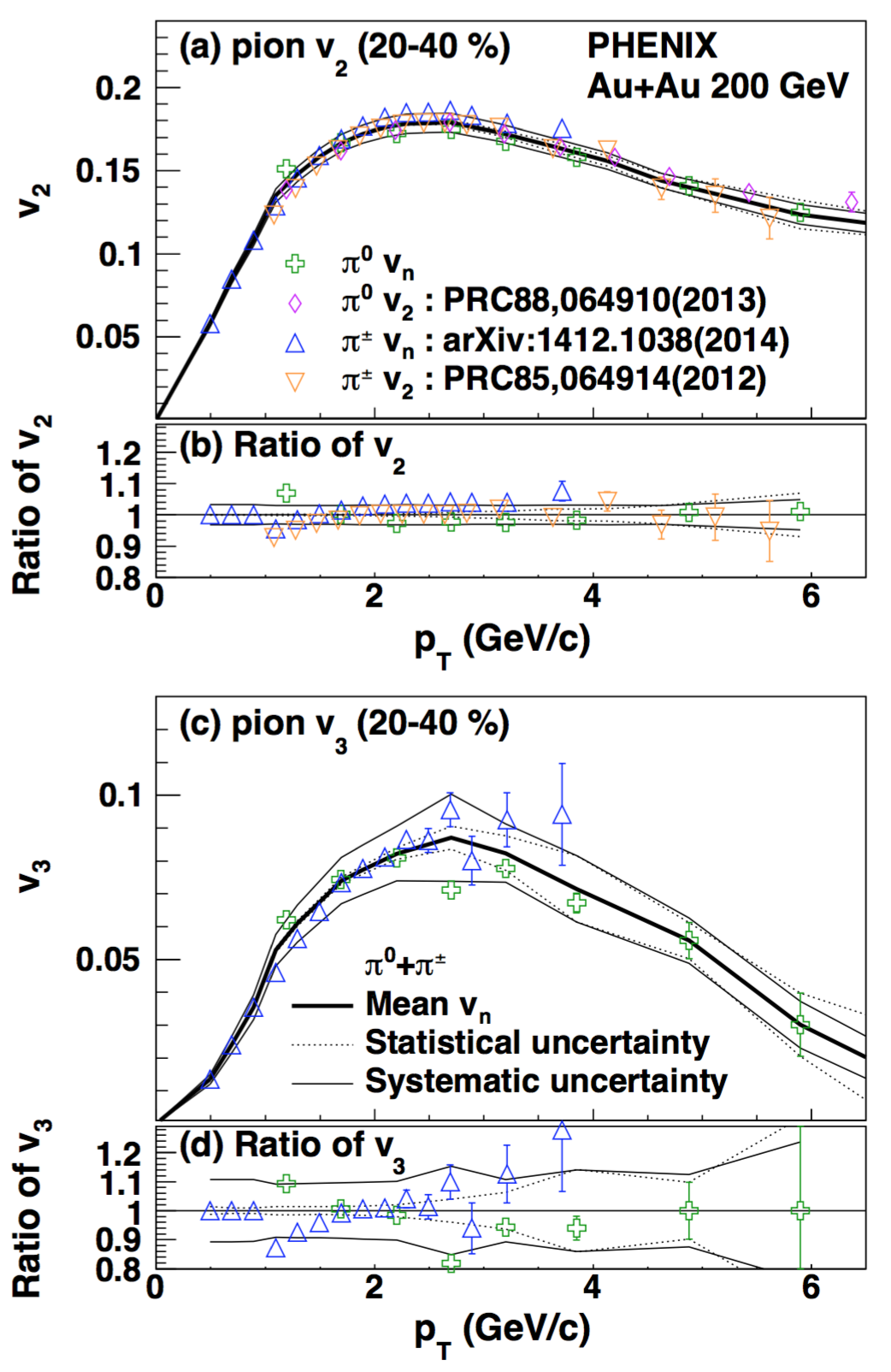}
		\caption{
        		Top panels: charged and neutral pion $v_{2}$ (a) and $v_{3}$ (c) for the 20-40$\%$ centrality class, including previously published results. The averaged values used in our analysis are shown as a thick solid line together with the estimated statistical (dotted line) and systematic (light solid line) uncertainties. Bottom panels: ratio of the measured v2 (b) and v3 (d) values to the averaged values.}
		\label{fig:vn_dec}
\end{figure}

The decay photon  $v_{2}^{dec}$ and $v_{3}^{dec}$ are estimated using the measured yield and anisotropy of charged and neutral pion~\cite{adare2013azimuthal}~\cite{adare2016measurement}~\cite{adare2012deviation}, as shown in Fig.~\ref{fig:vn_dec}.

$v_{n}$ for heavier mesons is derived from  $v_{n}$ of pions by scaling with the kinetic energy:
\begin{align}
v_{n}^{meson}(KE_{T}) = v_{n}^{\pi}(KE_{T})
\end{align}
where $KE_{T} = m_{T}-m = \sqrt{p_{T}^{2}+m^{2}}-m$, $m$ is the mass of the corresponding meson.
The meson yields, momentum spectra and $v_{n}$ are the input of a Monte Carlo simulation and decayed through all decay channels including photons. The decay photon $v_{n}^{dec}$ is then calculated from the simulation result.

\begin{figure}[htb] 
        \centering \includegraphics[width=1.0\columnwidth]{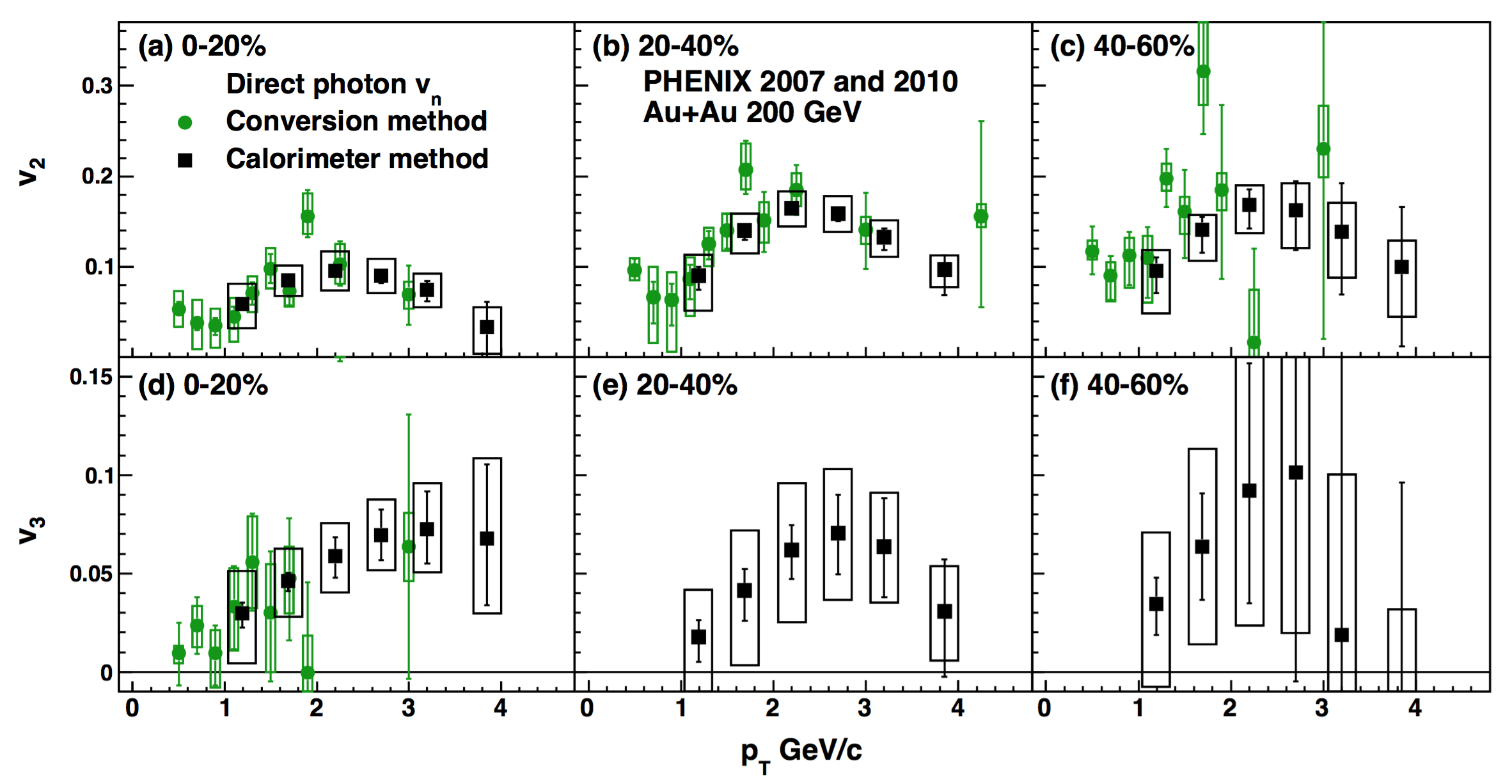}
		\caption{
        		Direct photon $v_{2}$ and $v_{3}$ at midrapidity ($|\eta|<0.35$) for Au+Au at 200GeV in centrality bins 0-20$\%$ (a, d), 20-40$\%$ (b, e), 40-60$\%$ (c, f). The green solid circles correspond to the conversion method, the black solid squares correspond to calorimeter method.}
		\label{fig:vn_dir}
\end{figure}

Together with the ratio of inclusive to decay photon yield $R_{\gamma}$ measured in \cite{adare2015centrality}, the direct photon $v_{2}^{dir}$ and $v_{3}^{dir}$ are obtained for three centralities using the two separate methods, with the results shown in Fig.~\ref{fig:vn_dir}.
As can be seen, the conversion method has a lower $p_{T}$ reach (down to 0.4\,GeV/$c$) than the calorimeter method, but the results are consistent in the overlap region. From Fig.~\ref{fig:vn_dir}, a large $v_{2}^{dir}$ is observed, which is comparable to the observed hadron $v_{2}$ at low $p_{T}$ side, and the $v_{2}^{dir}$ exhibits a strong centrality dependence. It also appears to converge toward zero at high $p_{T}$. A sizable $v_{3}^{dir}$ is observed as well, but there is no clear centrality dependence in this case.

\begin{figure}[htb] 
        \centering \includegraphics[width=0.9\columnwidth]{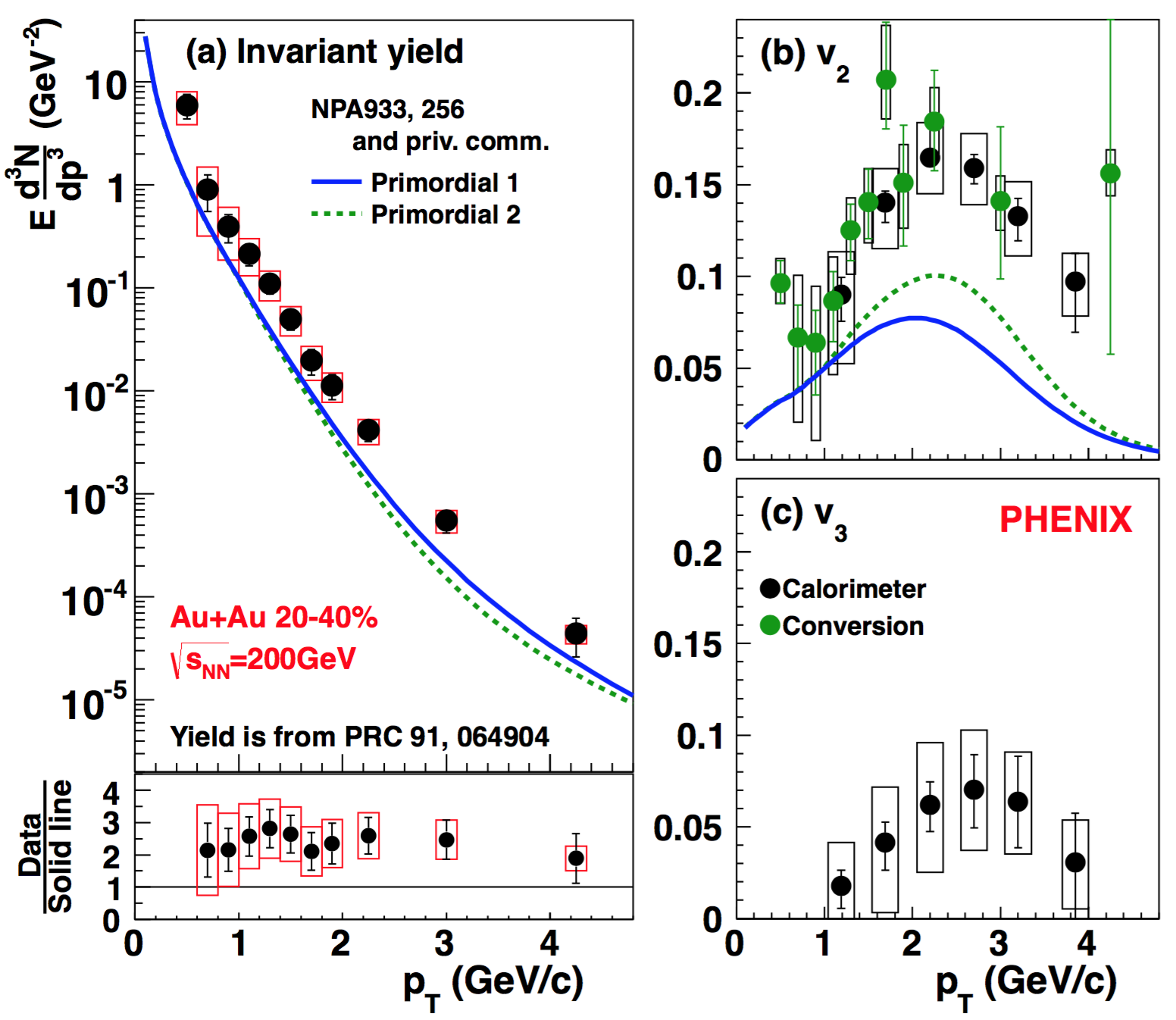}
		\caption{
        		Comparison of the direct photon yield and $v_{2}$ to the fireball model. The two curves for $v_{2}^{dir}$ correspond to two different parameterizations of the prompt photon component.}
		\label{fig:fire_ball}
\end{figure}

In Fig.~\ref{fig:fire_ball}, the data are compared with the ``fireball"
model~\cite{van2011thermal}. This scenario includes thermal photons from pQCD, QGP and HG with a fireball expansion profile. The calculation for $v_{2}^{dir}$ agrees with the data in the low $p_{T}$ region, but underestimates the yield in the entire $p_{T}$ range. The data are also compared with a semi-QGP and the PHSD models in~\cite{adare2016azimuthal}. None of those can describe both the large yield and large $v_{2}$ simultaneously to an adequate level. 

The results presented in this talk are using 2007 and 2010 datasets. PHENIX has also taken data for Au+Au collisions at 200\,GeV in 2014, which has much better statistics due to both the increased luminosity and more material for external conversions, because the HBD has been removed and a new detector (VTX) installed.  However, the previous technique to reconstruct external conversions at a fixed radius is no longer applicable. Therefore, we developed a new technique to reconstruct and identify external conversion photons without a priori knowledge of the conversion point. The new technique finds the true position where the external conversion takes place using all pairs of  $e^{\pm}$ tracks, based on the fact that the two tracks must have a zero opening angle at some common origin (the conversion point). The trajectories and momenta of $e^{\pm}$ are then reconstructed with respect to their true origin. Since there is no assumption about the track origin, the new technique can reconstruct conversions anywhere in the detector (the practical limits are 
5\,cm $<$ R $<$ 60\,cm). For the 2014 Au+Au dataset, we will focus mainly on the conversions at the third and fourth layer of VTX, which are located in between 10-25cm in radial direction from the collision vertex and correspond to $\sim10\%$ radiation length.  This new technique can also be applied to other datasets like Cu+Au, $^{3}$He+Au, p+p, p+Au and d+Au, taken since the VTX has been installed.

\section{Summary}
\label{}
In conclusion, PHENIX has measured the 2nd and 3rd Fourier coefficients of the azimuthal distributions of direct photons emitted at midrapidity for Au+Au collisions at $\sqrt{s_{NN}} = 200\,GeV$ using two independent analysis methods --- external conversion method and calorimeter method. The two results are consistent with each other in the overlap $p_{T}$ region, and the $v_{2}^{dir}$ result is consistent with the previously published result.

Large $v_{2}^{dir}$ and $v_{3}^{dir}$ are observed. The direct photon $v_{2}^{dir}$ is comparable to the hadron $v_{2}$ at $p_{T}<3GeV$, and $v_{3}^{dir}$ is similar to the hadron $v_{3}$ for the entire $p_{T}$ range.  While $v_{2}^{dir}$ exhibits a clear dependence on centrality, such dependence is not seen in the $v_{3}^{dir}$ result. We also compare the $v_{2}^{dir}$ and $v_{3}^{dir}$ results with several theoretical models. Those models fail to describe the yield and $v_{n}^{dir}$ simultaneously. As more new ideas are promoted to refine the theoretical picture, high quality measurement of yields and $v_{n}^{dir}$ from different colliding systems and energies with the newly developed technique will help to understand the direct photon production better and provide more constrains to the theoretical calculations. 





\begin{thebibliography}{00}



\bibitem{adare2010enhanced}
Adare, Ad, et al.``Enhanced production of direct photons in Au+ Au collisions at $\sqrt{s_{NN}} = 200$ GeV and implications for the initial temperature." Phys. Rev. Lett. 104.13 (2010): 132301.

\bibitem{adare2016azimuthal}
Adare, A., et al. ``Azimuthally anisotropic emission of low-momentum direct photons in Au + Au collisions at $\sqrt{{s}_{NN}}=200$ GeV" Phys. Rev. C 94.6 (2016): 064901.


\bibitem{adare2012observation}
Adare, A., et al. ``Observation of direct-photon collective flow in Au+ Au collisions at $\sqrt{s_{NN}} = 200$ GeV." Physical review letters 109.12 (2012): 122302.

\bibitem{adare2015centrality}
Adare, A., et al. ``Centrality dependence of low-momentum direct-photon production in Au+ Au collisions at $\sqrt{s_{NN}} = 200$ GeV." Phys. Rev. C 91.6 (2015): 064904.

\bibitem{stankus2011reaction}
Stankus, Paul W., Milan Matos, and Collaboration PHENIX. ``A reaction plane detector for PHENIX at RHIC." Nuclear Instruments and Methods in Physics Research Section A: Accelerators, Spectrometers, Detectors and Associated Equipment 636.1 (2011).

\bibitem{adare2013azimuthal}
Adare, A., et al. ``Azimuthal anisotropy of $\pi^{0}$ and $\eta$ mesons in Au+ Au collisions at $\sqrt{s_{NN}} = 200$ GeV." Physical Review C 88.6 (2013): 064910.

\bibitem{adare2016measurement}
Adare, A., et al. ``Measurement of the higher-order anisotropic flow coefficients for identified hadrons in Au+ Au collisions at $\sqrt{s_{NN}} = 200$ GeV." Phys. Rev. C 93.5 (2016): 051902.

\bibitem{adare2012deviation}
Adare, A., et al. ``Deviation from quark number scaling of the anisotropy parameter v 2 of pions, kaons, and protons in Au+ Au collisions at $\sqrt{s_{NN}} = 200$ GeV." Phys. Rev. C 85.6 (2012): 064914.

\bibitem{van2011thermal}
van Hees, Hendrik, Charles Gale, and Ralf Rapp. ``Thermal photons and collective flow at energies available at the BNL Relativistic Heavy-Ion Collider." Phys. Rev. C 84.5 (2011): 054906.


\end{thebibliography}



\end{document}